\documentclass[journal=ancac3,manuscript=article]{achemso}

\usepackage[version=3]{mhchem} % Formula subscripts using \ce{}
\usepackage{color}
\usepackage{gensymb}
\usepackage{hyperref}
\usepackage{pdfpages}
\usepackage{xcolor}
\usepackage{url}
\usepackage{hyperref}
\usepackage[utf8]{inputenc}  % If using pdfLaTeX
\usepackage[T1]{fontenc}
\usepackage{lmodern}
\DeclareUnicodeCharacter{0301}{\'{ }}
\newcommand{\commt}[1]{{\color{black}{#1}}}

\author{Ricardo Javier Peña Román}
\email{ricardo.pena_roman@tu-dresden.de}
\affiliation[MPI]
{Max Planck Institute for Solid State Research, 70569 Stuttgart, Germany}
\alsoaffiliation[TUD]
{Institute of Solid State and Materials Physics, Dresden University of Technology, 01069 Dresden, Germany}
\alsoaffiliation[cluster]
{Würzburg-Dresden Cluster of Excellence ctd.qmat, Dresden 01069, Germany}

\author{Sandip Maity}
\affiliation[MPI]
{Max Planck Institute for Solid State Research, 70569 Stuttgart, Germany}
\alsoaffiliation[TUD]
{Institute of Solid State and Materials Physics, Dresden University of Technology, 01069 Dresden, Germany}

\author{Fabian Samad}
\affiliation[Helm]
{Institute of Ion Beam Physics and Materials Research, Helmholtz-Zentrum Dresden-Rossendorf, 01328 Dresden, Germany}
\alsoaffiliation[TUC]
{Institute of Physics, Chemnitz University of Technology, 09107 Chemnitz, Germany}

\author{Dinesh Pinto}
\affiliation[MPI]
{Max Planck Institute for Solid State Research, 70569 Stuttgart, Germany}
\alsoaffiliation[EPFL]
{Institute de Physique, École Polytechnique Fédérale de Lausanne, CH-1015 Lausanne, Switzerland}

\author{Simon Josephy}
\author{Andrea Morales}
\affiliation[TUC] {QZabre AG, 8050 Zurich, Switzerland}
\author{Attila Kákay}
\affiliation[Helm]
{Institute of Ion Beam Physics and Materials Research, Helmholtz-Zentrum Dresden-Rossendorf, 01328 Dresden, Germany}
\author{Klaus Kern}
\affiliation[MPI]
{Max Planck Institute for Solid State Research, 70569 Stuttgart, Germany}
\alsoaffiliation[EPFL]
{Institute de Physique, École Polytechnique Fédérale de Lausanne, CH-1015 Lausanne, Switzerland}

\author{Olav Hellwig}
\affiliation[Helm]
{Institute of Ion Beam Physics and Materials Research, Helmholtz-Zentrum Dresden-Rossendorf, 01328 Dresden, Germany}
\alsoaffiliation[TUC]
{Institute of Physics, Chemnitz University of Technology, 09107 Chemnitz, Germany}
\author{Aparajita Singha}
\email{aparajita.singha@tu-dresden.de}
\affiliation[MPI]
{Max Planck Institute for Solid State Research, 70569 Stuttgart, Germany}
\alsoaffiliation[TUD]
{Institute of Solid State and Materials Physics, Dresden University of Technology, 01069 Dresden, Germany}
\alsoaffiliation[cluster]
{Würzburg-Dresden Cluster of Excellence ctd.qmat, Dresden 01069, Germany}

\title[An \textsf{achemso} demo]
{Nanoscale magnetometry of a synthetic three-dimensional spin texture}

\keywords{Synthetic antiferromagnet, Multilayered structures, 3D spin texture, Domain wall, NV-magnetometry, $T_1$ relaxometry, Magnetic noise.}

\begin{document}

\begin{abstract}
Multilayered synthetic antiferromagnets (SAFs) are artificial three-dimensional (3D) architectures engineered to create novel, complex, and stable spin textures. Non-invasive and quantitative nanoscale magnetic imaging of the two-dimensional stray field profile at the sample surface is essential for understanding the  fundamental properties of the spin-structure and being able to tailor them to achieve new functionalities. However, the deterministic detection of spin textures and their quantitative characterization at the nanoscale remain challenging. Here, we use nitrogen-vacancy scanning probe microscopy (NV-SPM) under ambient conditions to perform \commt{the first quantitative vector-field magnetometry measurements in the multilayered SAF [(Co/Pt)$_{5}$/Co/Ru]$_{3}$/(Co/Pt)$_{6}$}. We investigate the static and dynamic nanoscale properties of antiferromagnetic domains \commt{with boundaries hosting "one-dimensional" ferromagnetic stripes with $\sim$ 100 nm of width and periodic modulation of the magnetization. By employing NV-SPM measurements in different imaging modes and involving NV-probes with various crystallographic orientations, we demonstrate distinct fingerprints emerging from GHz-range spin noise and constant stray fields on the order of several mT}. This provides quantitative insights into the structure of domains and domain walls, as well as into magnetic noise associated with thermal spin-waves. Our work opens up new opportunities for quantitative vector-field magnetometry of modern magnetic materials with tailored 3D spin textures and stray field profiles, and potentially novel spin-wave dispersions—in a quantitative and non-invasive manner, with exceptional magnetic sensitivity and nanometer scale spatial resolution.

\end{abstract}

\section{Introduction}

Synthetic antiferromagnets (SAFs) are artificial magnetic systems that offer unique opportunities for fundamental research and applications due to their highly tunable static and dynamic properties \cite{Duine2018,Wang2023,Wang2024-bk,Xu2021,Rong2024}. In particular, SAFs present stable magnetic textures and high-frequency spin dynamics, which are ideal properties for magnetic storage technologies and spintronic devices \cite{Legrand2020,Pham2024,Godinho2024,Chen2021,Lonsky2022}.
\commt{Multilayered SAFs are systems with a more complex layered structure, designed to tune magnetic energy terms to create novel magnetic textures.
\cite{Hellwig2003,HELLWIG2007,Böhm2022,Mouhoub2023,Wang2023}. For instance, systems with perpendicular magnetic anisotropy (PMA) can be engineered to stabilize stripe and bubble domain states \cite{Kiselev2010}, to induce a domain wall (DW) transition from antiferromagnetic (AF) DWs to DWs with ferromagnetic (FM) cores\cite{Hellwig2003-prl}, or to stabilize FM domains separated by AF DWs when the interlayer exchange energy is reduced below the demagnetization energy \cite{Salikhov2025}}. From a dynamic point of view, dipolar interaction can affect the spin-wave modes and induce nonreciprocity on the spin-wave dispersion \commt{\cite{Gallardo2019,Jimenez-Bustamante2025-ub}}, giving rise to complex spin-wave propagation. \commt{Additionally, multilayered structures with periodic magnetic stripe domain patterns exhibit spin-wave modes and dispersion relations analogous to those of a one-dimensional magnonic crystal, which can be used to control spin-wave propagation.\cite{Banerjee2017,GRUSZECKI202129}.} Therefore, these types of multilayered architectures are a \commt{powerful and application friendly approach} to develop new magnetic materials with spin textures and properties useful to improve the performance of SAFs and to design the next generation of magnetic devices \cite{Gubbiotti_2025,Zhang2021,Lavrijsen2013, Fernandey2016}.

The direct imaging of the two-dimensional stray magnetic field profile at the sample surface with nanoscale spatial resolution can provide insights \commt{into the properties of individual domains and DWs}, which is a crucial step toward understanding and improving the fundamental properties and performance of multilayered SAFs. Magnetic force microscopy (MFM) combined \commt{with macroscopic magnetometry} have been employed to investigate multilayered SAFs and to connect the nanoscale magnetism of every domain/DW with the macroscopically observed properties, such as magnetization or magnetic hysteresis  \cite{Hellwig2003,Hellwig2003-prl,HELLWIG2007,Salikhov2025,Böhm2022,Koch2020,Salikhov2021,Salikhov2022}. However, the \commt{field} of a few tens of mT of the MFM probe  \cite{Balasubramanian2008-ec} can induce \commt{reversible changes or irreversible} reorientation of the magnetic moments while scanning the sample \cite{Stepanova2022,Zelent2021,Casiraghi2019-wd}, hindering access to the intrinsic \commt{undisturbed} magnetic texture. Furthermore, macroscopic magnetometry provides only information about the average behavior of a large number of magnetic domains. Moreover, \commt{many} magnetic imaging techniques probe only static properties non-quantitatively and lack the sensitivity required to investigate phenomena with weak stray magnetic field (such as AF order). Therefore, despite the enormous effort in the last few years, there is still a remaining challenge regarding the deterministic detection of magnetic textures in multilayered SAFs, along with the local, quantitative, and non-invasive characterization of both static and dynamic properties. 

Nitrogen-Vacancy scanning probe microscopy (NV-SPM) has recently gained immense interest as a robust quantum sensing tool. It utilizes the quantum properties of a single NV defect in diamond for quantitative and non-invasive magnetic imaging, 
where the magnetic vector-field sensitivity can be as high as nT with an exceptional nanoscale spatial %and nanoscale quantum sensing with high magnetic sensitivity on the order of few $\mu$T/$\sqrt{\text{Hz}} and$ 
resolution, in the range of tens of nanometers \cite{Degen2017,Finco2023,Rovny2024-bz,Budakian_2024,Christensen_2024,Degen2025}.  NV-SPM has been previously employed to investigate ultra-thin SAFs or FM thin films to determine the nature of DWs and strength of the Dzyaloshinskii-Moriya interaction (DMI) \cite{Tetienne2014,Tetienne2015-rc,Gross2016}, DW motion \cite{Tetienne-2014,McLaughlin2023}, spin-wave dispersion and magnetic noise \cite{Finco2021-ne,Finco2025}. In such systems, the magnetic thickness is typically less than 2 nm with a few mT of stray field, for which unwanted effects that often make quantitative measurements difficult, such as spin-mixing of the NV-spin states and suppression of the quantum sensor's signal due to high off-axis stray fields, are not present. Consequently, magnetic imaging of thick multilayered SAFs with complex and novel spin textures and dynamics, producing both a large stray field of tens of mT and spin noise, \commt{remained} unexplored using NV-SPM.

In this work, we use NV-SPM under ambient conditions to investigate the magnetic texture of \commt{the multilayered SAF of the type [(Co/Pt)$_{X-1}$/Co/Ru]$_{N-1}$/(Co/Pt)$_{X}$ with $X=6$, and $N=4$}. We explore the domains and DWs structure, as well as their static and dynamic properties, by performing measurements in different imaging modes and employing NV-probes with various crystallographic orientations to identify distinctive fingerprints arising from both constant stray magnetic fields and spin noise. We use optically detected magnetic resonance (ODMR) and $T_1$ relaxometry measurements to demonstrate the presence of both sample stray fields on the order of mT and spin noise in the GHz range. We perform qualitative magnetic imaging with commercial diamond probes containing a single NV and fabricated from (100)- and (110)-oriented diamond crystals\cite{Welter_2022}.\commt{ By using a (111)-oriented single NV probe\cite{Rohner_2019}, we quantitatively resolve the nanoscale DW structure and the spatial distribution of the sample stray field}. Combining this with micromagnetic simulation,  we \commt{confirm} a three-dimensional (3D) model of the spin texture that agrees well with our quantitative measurements. Our work opens novel opportunities to understand and characterize static and dynamic properties in layered architectures with complex magnetic textures and spin-waves, quantitatively and non-invasively, with exceptionally high magnetic sensitivity and nanometer-scale spatial resolution.

\section{Results and discussions}

\subsection{NV-SPM imaging in a multilayered SAF}

\begin{figure}[H]
    \centering
\includegraphics[width=6.4 true in]{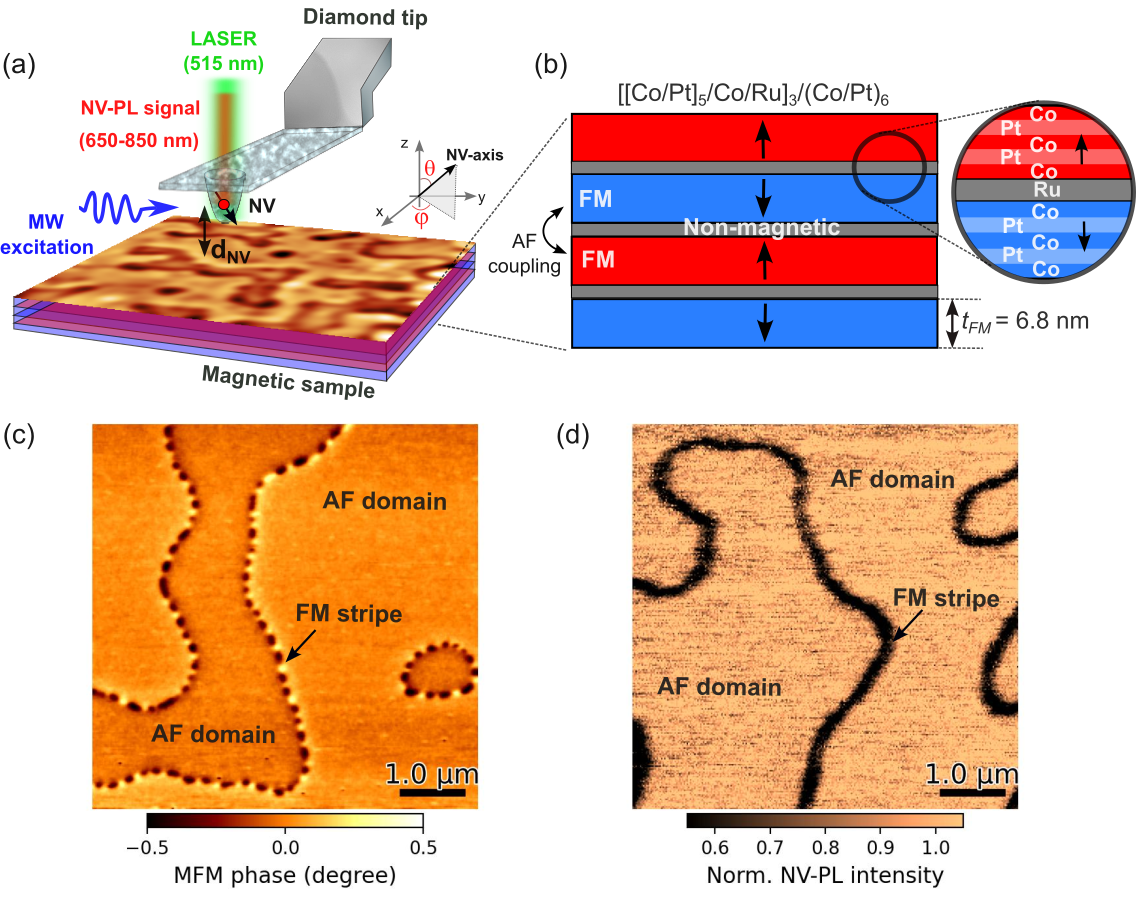}
\caption{\label{Fig:NV-sample} \textbf{Magnetic imaging of a multilayered synthetic antiferromagnet.} (a) A single NV center in a scanning diamond probe is used for sensing the magnetic field and noise of the sample at a given NV-sample distance $d_{\text{NV}}$.  The NV spin state is optically initialized with a green laser (515 nm) and read out by detecting the NV-PL signal (650-850 nm) after interaction with the sample. \commt{Qualitative magnetic imaging is possible via MW-free NV-PL maps with and without applying external magnetic field}. For ODMR measurements, MW excitation in the GHz range is required to promote the electron spin transition. The sample stray field and noise are sensed along and perpendicular to the NV-axis, respectively. The angles $(\theta_{\text{NV}},\varphi_{\text{NV}})$ define the orientation of the NV-axis in the laboratory frame. (b) Schematic structure of the multilayered SAF \commt{[[Co(0.55 nm)/Pt(0.70 nm)]$_{5}$/Co(0.55 nm)/Ru(0.75 nm)]$_{3}$/[Co(0.55 nm)/Pt(0.70 nm)]$_{6}$}. (c)  6 $\mu$m x 6 $\mu$m MFM image of the multilayered SAF. (d)  6 $\mu$m x 6 $\mu$m NV-PL quenching map recorded with a (100)-oriented NV probe and \commt{without applying external magnetic field}, pixel size 24 nm/px and acquisition time 24 ms/px. The intensity is normalized with respect to the average counts on the AF domains (bright areas).}
\end{figure}

As illustrated in Figure \ref{Fig:NV-sample}(a), NV-SPM employs a single NV defect in a diamond scanning probe (see section 1 in the Supporting Information (SI)). The NV sensor is used to map the sample surface and spatially resolve the magnetic texture. Due to its spin-dependent photoluminescence (PL), the NV-spin state can be optically initialized with a green laser (515 nm) and read out by detecting the red NV-PL signal in the wavelength range of 650-850 nm (see methods). Qualitative magnetic imaging can be performed by mapping the intensity of the NV-PL signal while scanning the sample.  For direct quantification of the sample stray field by ODMR, microwave (MW) excitation is applied  to promote the NV spin transition. Thus, the stray magnetic field of the sample is determined from the Zeeman splitting  produced by its component along the NV-axis, which is oriented along the crystallographic direction ⟨111⟩ in the diamond probe. The orientation of the NV-axis is defined by the angles $(\theta_{\text{NV}},\varphi_{\text{NV}})$ in the laboratory frame. From our ODMR measurements under ambient conditions, we estimate a DC magnetic field sensitivity of $ \sim$ 2 $\mu$T/$\sqrt{\text{Hz}}$. The value of the detected stray field and the spatial resolution for magnetic imaging depend on the NV-sample distance, which is $d_{\text{NV}}=$ 36-76  nm for all our measurements (see Section 2 in SI).

Figure \ref{Fig:NV-sample}(b) illustrates the structure of the multilayered SAF investigated in this work. \commt{The stacking of 6 Co layers with 5 Pt interlayers} give rise to FM building blocks with PMA. i.e. out-of-plane (OOP) magnetization parallel or antiparallel to the film normal axis. The \commt{insertion of a non-magnetic material like Ru introduces an AF-coupling between adjacent FM blocks through the Ruderman–Kittel–Kasuya–Yosida exchange interaction. The strength of the AF-exchange interaction can be tuned via the Ru thickness.  \cite{Duine2018,Wang2023,HELLWIG2007}.}  In our sample, we have the AF-coupling of four FM blocks, and the full structure  is described as \commt{[[Co(0.55 nm)/Pt(0.70 nm)]$_{5}$/Co(0.55 nm)/Ru(0.75 nm)]$_{3}$/[Co(0.55 nm)/Pt(0.70 nm)]$_{6}$}. The numbers in nm correspond to the thickness of every thin film along the stacking. Thus the magnetic thickness of a single FM block is  $t_{FM}= 5\times[0.55\text{ nm}(\text{Co})+0.70\text{ nm}(\text{Pt})] + 0.55\text{ nm}(\text{Co})=  6.8$ nm. \commt{The hysteresis} curve measured on this sample shows four discrete steps (see section 3 in SI), where every step corresponds to the magnetic reversal of \commt{one} of the four FM blocks, with an average saturation magnetization \commt{$M_s=(833\pm 83)$ kA/m} for configurations with a full magnetization pointing up/down. 

MFM images recorded after sample saturation with an in-plane applied field reveal a remanent state constituted by \commt{one-dimensional FM stripe domains with periodic bright and dark contrast at the boundary between AF domains\cite{Hellwig2003-prl,HELLWIG2007}}, as shown in Figure \ref{Fig:NV-sample}(c). The \commt{AF domain} sizes are primarily in the micrometer range, with $\sim$ 100 nm-wide
FM stripes and average periodicity of the contrast of $\sim$ 300 nm (see section 3 in SI) . The width of the stripes and the periodicity of the contrast can vary slightly from one stripe to another, but typically remain on the order of hundreds of nanometers. The alternating bright and dark contrast observed along the FM stripe, as well as, the slight difference in magnetic contrast between adjacent AF domains, originates from the parallel and antiparallel alignment \commt{of} the sample stray field and the magnetic MFM tip. Such magnetic texture has been observed by Hellwig \emph{et al.} in similar samples with \commt{various $X$, $N$, and layer thicknesses
%$X = 6, 7, 8$ and $N = 4, 12, 17, 18$ 
\cite{Hellwig2003-prl,HELLWIG2007}.
However, gaining further quantitative microscopic details on the spin texture of domains and DWs along with their dynamic properties are beyond the scope of standard MFM measurements. }

To this end, we employ the NV-SPM approach. Figure \ref{Fig:NV-sample}(d) shows an image acquired in the NV-PL quenching mode with a (100)-oriented NV probe. The NV-PL signal is detected pixel by pixel as the sample is scanned across the NV tip which is kept stationary at the focus of the initializing green laser beam. MW excitation is not applied during the measurements in this imaging mode. We observe large bright areas on the AF domains. \commt{Along the FM stripe domains} the PL signal is quenched by about 40\%. The NV-PL signal can be quenched either due to spin-mixing effects induced by strong off-axis stray fields \cite{Tetienne_2012} or due to a decrease of the spin-relaxation time ($T_1$) in case spin noise is present with frequencies resonant with the NV spin transition \cite{Finco2021-ne,Rollo2021}. Contrary to MFM measurements, there is no periodic variation of the contrast. \commt{However, MFM-like contrast can be achieved in NV-PL maps only after applying a suitable off-axis (not aligned with the NV-axis) external magnetic field, as described in the following section.}

\subsection{MFM-like contrast in NV-PL quenching maps}

\begin{figure}[H]
\centering
\includegraphics[width=6.5 true in]{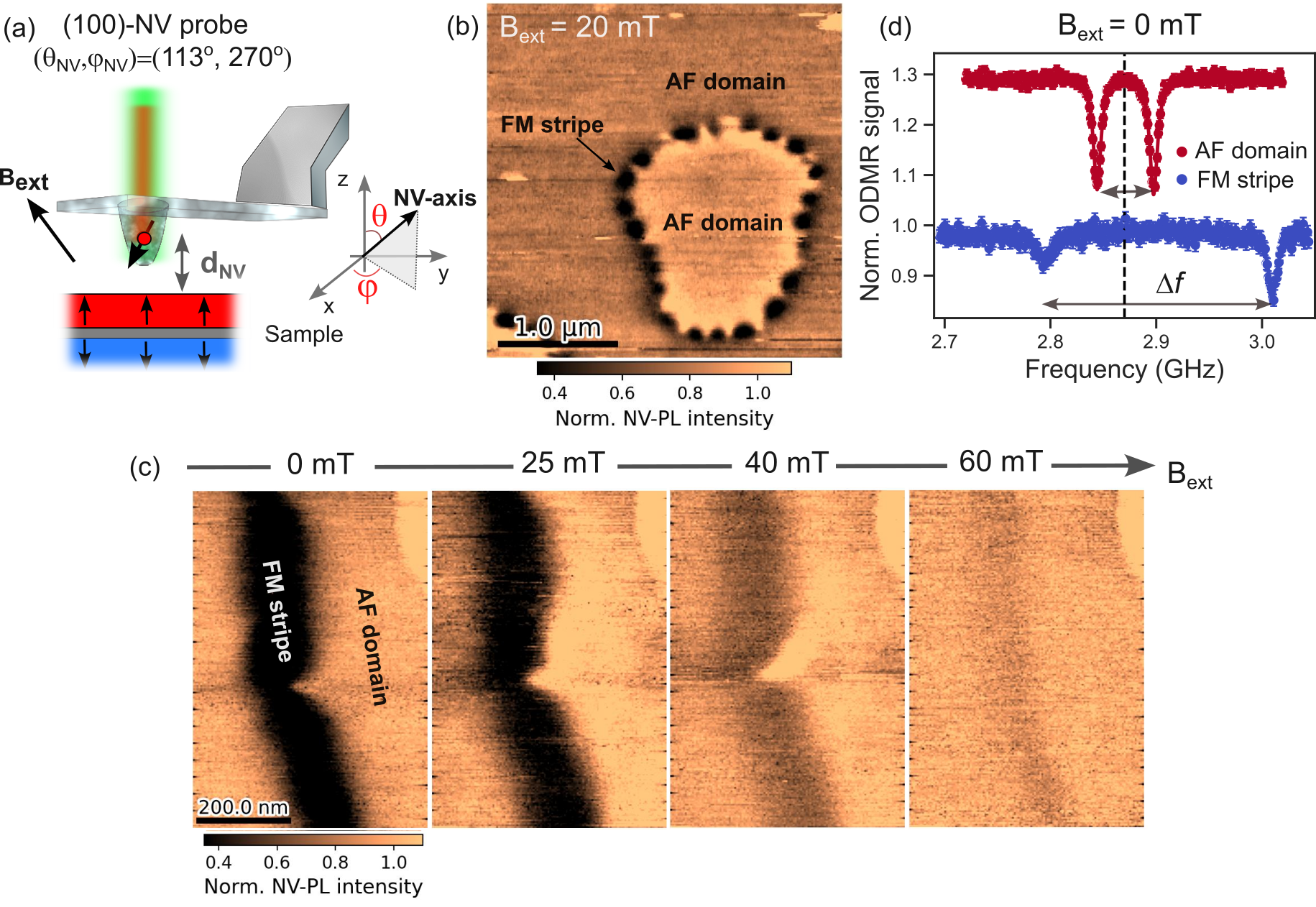}
\caption{\label{Fig:PL_100}  \textbf{Effects of an \commt{external} off-axis magnetic field on the contrast of NV-PL maps.} (a) Schematic illustration of the relative orientation between the (100)-oriented NV probe and the sample magnetization during NV-SPM measurements. The orientation of the NV-axis is determined by the coordinates $(\theta_{\text{NV}},\varphi_{\text{NV}}) = (113\degree,270\degree )$  in the laboratory frame. (b) 3 $\mu$m x 3 $\mu$m NV-PL quenching map recorded in presence of an external off-axis field  $B_{\text{ext}}$ = 20 mT applied along $(\theta_{\text{ext}},\varphi_{\text{ext}}) = (54.8\degree,180\degree )$. Pixel size  100 nm/px and acquisition time 25 ms/px. (c) High resolution NV-PL quenching maps (500 nm x 720 nm) on a FM DW recorded at different value for the external field. Pixel size 4 nm/px and acquisition time 25 ms/px. (d) Normalized and vertically shifted ODMR spectra measured on the AF domain and FM DW without external field.}
\end{figure}

We determined $(\theta_{\text{NV}},\varphi_{\text{NV}}) = (113\degree,270\degree )$ for the (100)-oriented NV probe (see section 2 in SI).  Figure \ref{Fig:PL_100}(a) illustrates the relative orientation between the single NV-spin and the magnetization of the sample. We observe a different contrast between the AF domains and \commt{along the FM stripes}, when an external field off-axis is applied during the acquisition of NV-PL images, as can be seen in Figure \ref{Fig:PL_100}(b) for $B_{\text{ext}}= 20$ mT.  \commt{The external field is applied along $(\theta_{\text{ext}},\varphi_{\text{ext}}) = (54.8\degree,180\degree )$, with an angle $\alpha \approx103 \degree$ with respect to the NV-axis}. The image is similar to the one obtained with MFM, but without the inherent local effect of the field from the tip; here, the external field is globally and uniformly applied to both the sample and the tip. High-resolution images of \commt{a FM stripe between AF domains} are shown in Figure \ref{Fig:PL_100}(c). The contrast is modified in the NV-PL quenching images when the intensity of the applied field is increased from 0 mT to 60 mT due to field compensation  or competition between the external and sample stray field.  

\commt{To understand the origin of the NV-PL quenching and explain the effect of the external field on the contrast of the NV-PL maps, we first performed} local ODMR measurements without any external field.  Figure \ref{Fig:PL_100}(d) shows ODMR spectra measured locally when the tip is on an AF domain and on a \commt{FM stripe domain} (see section 4 in SI for additional measurements). On the AF domain, we observe an ODMR spectrum with around 20 \%  of contrast, and the splitting $\Delta f$ is symmetric with respect to the zero-field splitting (ZFS) frequency $f_0=2.87$ GHz indicated by the vertical dashed line. However, in contrast, the splitting $\Delta f$ is asymmetric with respect to $f_0$ on the \commt{FM stripe} and the contrast also decreases to less than 10\% for the dip at 2.80 GHz. The features observed on the \commt{FM stripe} are indications  that the NV-PL quenching is a consequence of spin-mixing effects \cite{Tetienne_2012} produced by  components of several mT of the sample stray magnetic field that are perpendicular to the NV-axis. At some locations along the \commt{FM stripe}, the stray field can be too strong to perform any ODMR measurements (see section 4 in SI). When spin-mixing effects are present, the NV-PL intensity is reduced since there are NV-spin transitions involving states that are linear combinations of dark ($m_s=\pm1$) and bright ($m_s=0$) states. NV-PL quenching is expected along the \commt{boundary between AF domains} as these are the regions in the sample where we \commt{have} FM order with a strong field. 

To further explain the effect of the external off-axis field and the MFM-like contrast in NV-SPM, we use the model proposed in Section 5 of the SI. We consider all the field components perpendicular to the NV-axis that could produce quenching of the PL signal due to spin-mixing. The net magnetic field perpendicular to the NV-axis has contributions from the external off-axis field ($\boldsymbol{B}_{\text{ext}}$) and the sample stray magnetic field  ($\boldsymbol{B}_{\text{s}}$): 

\begin{equation}\label{1}
\boldsymbol{B}_{\text{total}}^{(\perp, \text{NV})} = \boldsymbol{B}_{\text{ext}}^{(\perp, \text{NV})} + \boldsymbol{B}_{s}^{(\perp,\text{NV})}, 
\end{equation}
where :
\begin{equation}\label{2}
{B}_i^{(\perp, \text{NV})} = {B}_i \sqrt{1 - \left[ \sin \theta_i \sin \theta_{\text{NV}} \cos (\varphi_i - \varphi_{\text{NV}}) + \cos \theta_i \cos \theta_{\text{NV}} \right]^2}.
\end{equation}

 with  $i=\text{ext}, \text{ or }\text{ s}$. To simplify the description, we consider only the  components of the stray field of the sample that are parallel and antiparallel to the sample surface. That is, components with $(\theta_{\text{s}},\varphi_{\text{s}}) = (0\degree,0\degree )$ and $(\theta_{\text{s}},\varphi_{\text{s}}) = (180\degree,0\degree )$, for OOP magnetization pointing up and down, respectively. The angle between these OOP stray field components and the NV-axis are $113 \degree$ and  $67 \degree$, respectively. Due to the magnetic field compensation, different degrees of quenching are expected for regions where $ B_{\text{ext}}^{(\perp, \text{NV})}$ and $ B_{\text{s}}^{(\perp, \text{NV})}$ are parallel or antiparallel. For example, considering the values $B_{{s}} = \pm5$ mT for the stray field on the \commt{FM stripe}  (which is enough to produce spin-mixing in the case of a large-angle off-axis field\cite{Tetienne_2012})  and $B_{\textbf{ext}} = 20$ mT, from equation (\ref{1}) and (\ref{2}) we can obtain $B_{\text{total }\uparrow}^{(\perp, \text{NV})} = 24$ mT and $B_{\text{total }\downarrow}^{(\perp, \text{NV})} = 15$ mT for regions where the sample stray field components are pointing up and down, respectively. We assume that the relative degree of quenching produced by these two fields in the NV-PL signal is proportional to the difference normalized by the total field. i. e. $|B_{\text{total}\uparrow}^{(\perp, \text{NV})} - B_{\text{total}\downarrow}^{(\perp, \text{NV})} |/[{B_{\text{total}\uparrow}^{(\perp, \text{NV})} + B_{\text{total}\downarrow}^{\perp, \text{NV})} }] =0.23$. We interpret  this as a relative contrast of $C=23\%$ between the dark and the bright areas in the NV-PL maps,  leading to the MFM-like contrast in Figure \ref{Fig:PL_100}(b). This relative contrast decreases when increasing the intensity of the external field (see section 5 in SI), being $C=7\%$ for $B_{\text{ext}} = 60$ mT. This explains the almost vanishing contrast on \commt{the FM stripe} in Figure \ref{Fig:PL_100}(c). \commt{It is noteworthy that the effect is reversible. Once the external field is switched off, a PL map similar to that for $B_{\text{ext}}=0$ mT is recovered}. For $B_{\text{ext}}=0$ mT, the contribution is from the sample, i.e., $B_s = \pm5$ mT, which produces the same degree of quenching. In general, $B_s$ is not fully $\pm$5 mT; it can be smaller or larger. However, for spin-mixing, the difference in the PL drop produced by a variation in $B_s$ of 1-5 mT would be around 5\% \cite{Tetienne_2012, Rondin_2014}. Thus, it would not be possible to differentiate between the PL quenching produced by two close values of the stray field with a significant contrast. Therefore, the contrast along the FM stripe and within the AF domains will be uniform for domains with magnetization pointing upward and downward, in the absence of an off-axis $B_{\text{ext}}$, as also observed in Figure \ref{Fig:NV-sample}(d). Additional NV-PL maps with MFM-like contrast in a different sample are shown in section 6 of the SI.
 
\subsection{Internal structure of the ferromagnetic \commt{stripe}}

\begin{figure}[H]
    \centering
\includegraphics[width=6.5 true in]{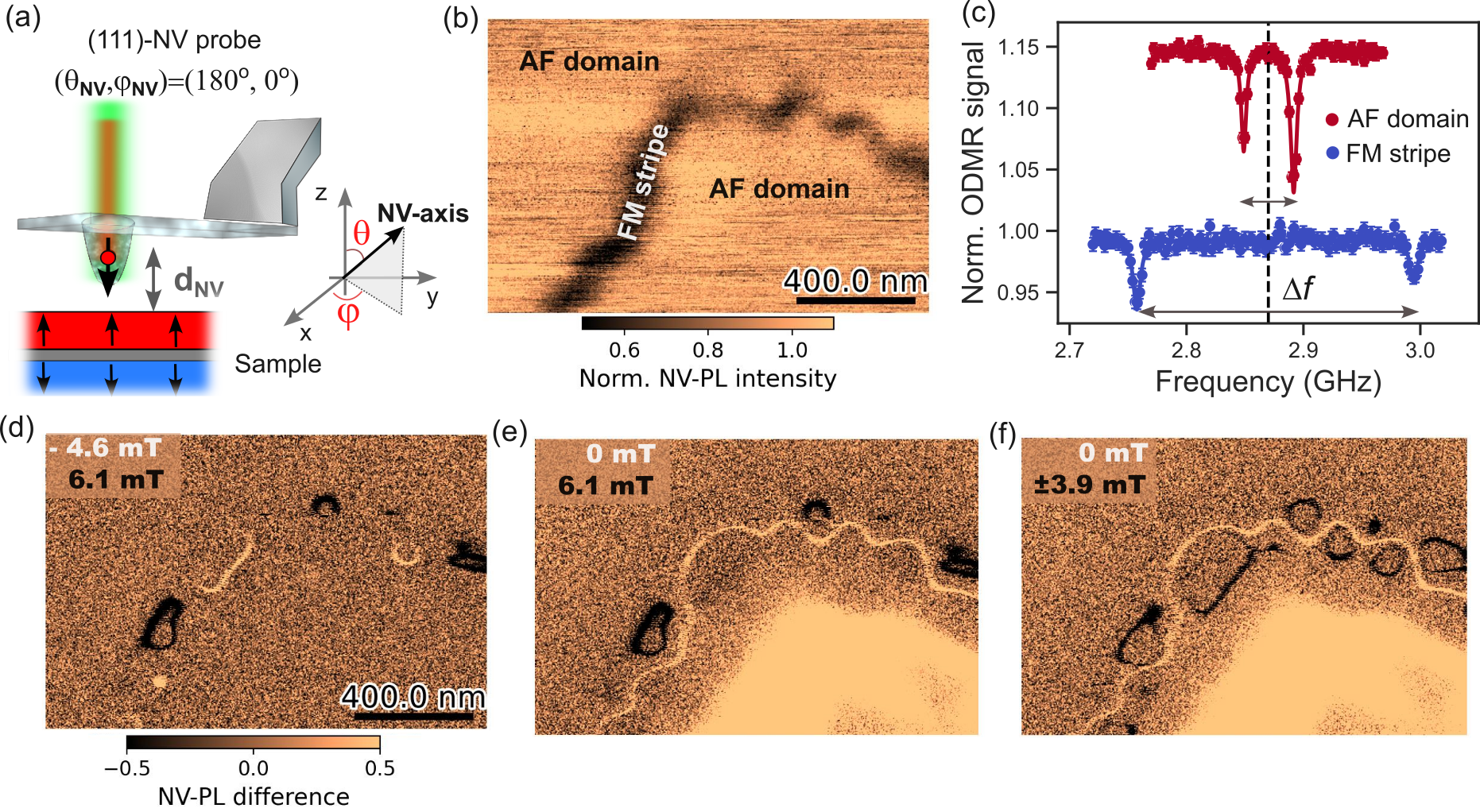}
\caption{\label{Fig:PL_111} \textbf{Resolving the internal structure of the FM stripe.} (a) Schematic illustration of the relative orientation between the NV-axis of the (111)-oriented NV probe and sample magnetization during NV-SPM measurements. The orientation of the NV-axis is determined by the coordinates $(\theta_{\text{NV}},\varphi_{\text{NV}}) = (180\degree,0\degree )$  in the laboratory frame. (b) NV-PL quenching map recorded without external field. (c) Normalized and vertically shifted ODMR spectra measured on the AF domain and FM DW. (d)-(f) Dual Iso-B contour images. The size, pixel size and acquisition time for all the images shown in this figure are 1.5 $\mu$m x 1.0 $\mu$m, 4 nm/px and 20 ms/px, respectively.}
\end{figure}

To avoid having strong stray field components producing spin-mixing and be able to perform quantitative measurements, we employed a (111)-oriented NV probe. With this diamond probe we determined $(\theta_{\text{NV}},\varphi_{\text{NV}}) = (180\degree,0\degree )$ (see section 2 in SI). Therefore, we can assume a collinear alignment between the NV-axis and the sample magnetization, as illustrated in Figure \ref{Fig:PL_111}(a). An NV-PL quenching map recorded without applying an external field is shown in Figure \ref{Fig:PL_111}(b). The PL signal is quenched only along the \commt{FM stripe between AF domains}. However, in contrast to the case of  (100)-NV tip, the local ODMR measurements in Figure \ref{Fig:PL_111}(c) show symmetric Zeeman splitting $\Delta f$ with respect to $f_0$, \commt{on both the AF domain and the FM stripe} (see Section 7 in SI for additional measurements). Based on these ODMR results, we can rule out a PL-quenching induce by spin-mixing. Thus, we attribute the NV-PL quenching to magnetic noise \cite{Finco2021-ne}. From the spectra depicted in Figure \ref{Fig:PL_111}(c), we can determine a magnetic field projected on the NV-axis of 0.6 mT and  4.0 mT \commt{on both the AF domain and the FM stripe}, respectively. 

\commt{To gain quantitative insight into the FM texture of the stripes observed between the AF domains in NV-PL quenching maps and MFM images, we first performed NV-SPM measurements in the iso-B imaging mode by acquiring NV-PL maps with a given MW frequency to resolve iso-magnetic field contours\cite{Rondin_2014} (see section 8 in SI). Dual iso-B images are obtained by subtracting the PL images at two fixed frequencies. In this case, we observe contour maps with two different values of the sample stray field.} The dual iso-B image in Figure \ref{Fig:PL_111}(d) reveals contours of FM domains in the nanometer range with a stray field  $B_{\text{NV}}=-4.6$ mT and  $+ 6.1$ mT. The orientation of the field changes periodically, in analogy with MFM. Zero-field regions are resolved in the dual iso-B image in Figure \ref{Fig:PL_111}(e). We observe an area of hundreds of nm$^2$ with  $B_{\text{NV}}= 0 $ mT at the AF domain on the right. Additionally, there is a zero-field line between domains. Since the NV-axis is essentially perpendicular to the sample surface, this zero-field line indicates that there is no field projection along the NV-axis. Therefore, it can be interpreted as the position at which the OOP component of the field is zero, which could define \commt{the position of the DW inside the FM stripe.  \cite{Tetienne2014,Tetienne2015-rc,McLaughlin2023}. FM contours with $B_{\text{NV}}=+ 6.1$ mT are observed along one side of the zero-field line, while domains with  opposite orientation of the field components are periodically separated by the zero-field line as shown in the image in Figure \ref{Fig:PL_111}(f) for FM domains with $B_{\text{NV}}= \pm 3.9$ mT.}

\subsection{Quantitative vector-field maps of a 3D spin texture}

\begin{figure}
\centering
\includegraphics[width=5.6 true in]{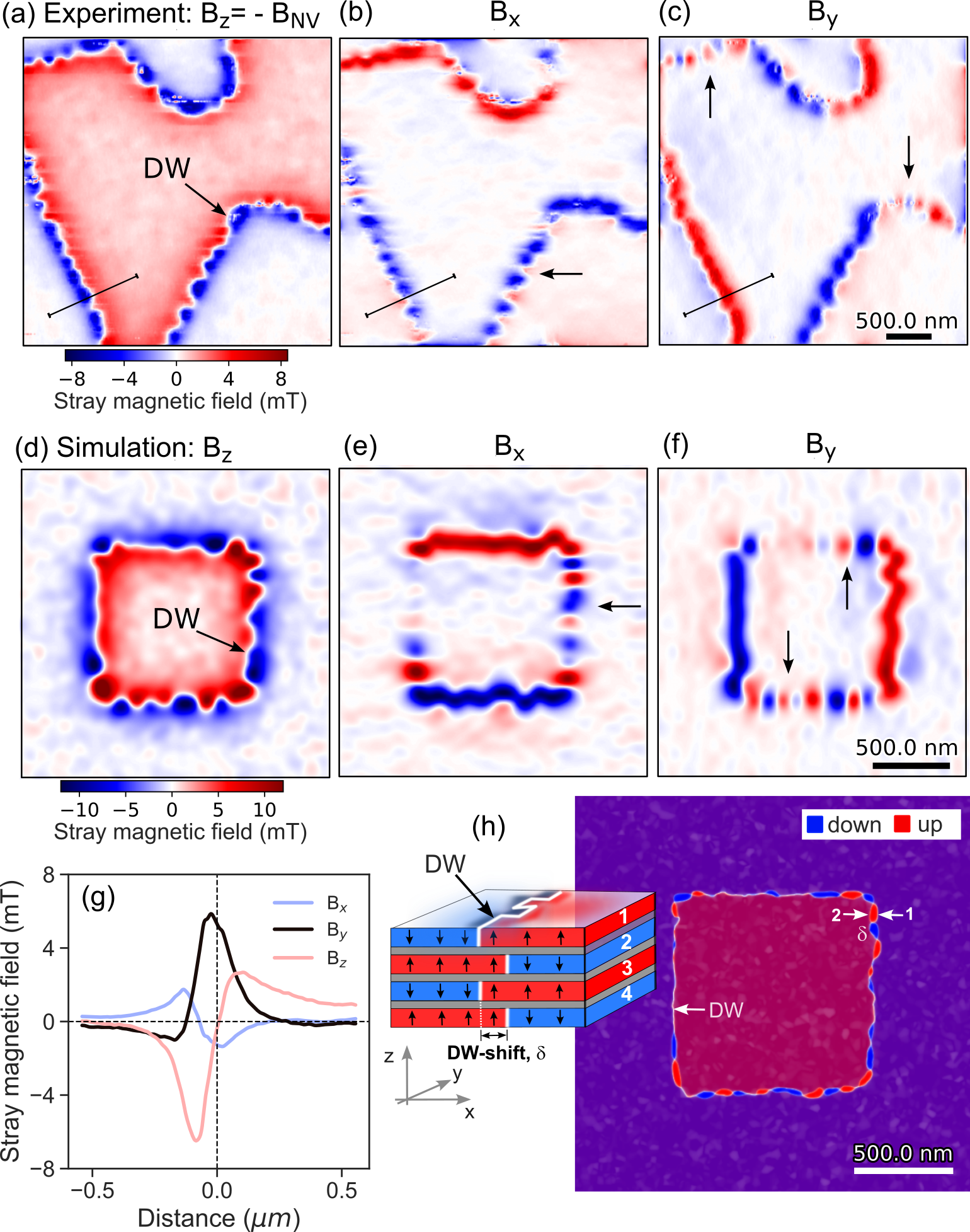}
\caption{\label{Fig:odmr}  \textbf{\commt{Stray field maps, DWs and FM cores}.} (a) Stray field map of the component projected along the NV-axis, $B_{z}= - B_{\text{NV}}$, pixel size of 3.5 nm/px and acquisition time of 100 ms/px. Reconstructed stray field map of the components along (b) the x-axis and (c) the y-axis. Micromagnetic simulated images of the stray field map produced by a square AF domain along (d) the z-axis, (e) the x-axis and (f) the y-axis. Field values are calculated at 75 nm from the sample surface. (g) Profiles of the OOP and IP components of the measured stray fields across the DW. (h) Overlap of \commt{the  simulated remanent OOP magnetization} of the top and second FM blocks to visualize the alternating left and right DW-shift that give rise to the FM cores around the DW with magnetic moments pointing up and down. The inset figure illustrates the 3D magnetic texture of the multilayered structure.}
\end{figure}

Furthermore, we record ODMR maps with a (111)-NV probe as shown in Figure \ref{Fig:odmr}(a). The field component projected along the NV-axis is spatially resolved over the sample surface. Given this crystallographic orientation of the tip, we are probing the $B_{z}$= - $B_{\text{NV}}$ component of the stray magnetic field. 2D maps of the in-plane (IP) stray field components can be reconstructed from the experimental map in Figure \ref{Fig:odmr}(a) by using the reverse propagation method in Fourier space \cite{Lima2009, dovzhenko2018} \commt{(see section 9 in SI)}. The obtained $B_{x}$ and $B_{y}$ stray field maps are depicted in Figure \ref{Fig:odmr}(b) and (c), respectively. The maximum value for the static stray field is $\approx \pm$ 8.0 mT \commt{at the boundary of the AF domains}. The intensity of the stray field progressively decreases when moving far away from the boundary by hundreds of nm. Both IP and OOP components of the stray field are in the range from hundreds of $\mu$T to 1mT on the AF domains. Even if we expect in a multilayered SAF structure with PMA a full compensation of the magnetization on the AF domains, there will be a small field sensed by the NV above the sample surface due to the different distances between the NV center and each of the FM blocks \cite{Finco2021-ne,Kiselev2008}. \commt{Additionally, from the images we observe a granular contrast, indicating magnetization variations (see section 10 in SI for a close-up ODMR map on the AF domains)}. These inhomogeneities or granularity are expected in sputtered multilayered samples with non-uniform thin film thicknesses or atomically flat surface that can induce local variation and non-compensation of the magnetization. 

Profiles taken across the region between the AF domains are plotted in Figure \ref{Fig:odmr}(g). For simplicity, we define the DW as the region where the OOP field component ($B_z$) is zero \cite{Tetienne2015-rc,McLaughlin2023, Tetienne2014} (centred at distance=0 in the profiles). Since it is the region where there is a sign reversal of the stray field  (as shown in the dual iso-B results). Note that it strictly corresponds to the actual DW center line ($m_z=0$, $z$-magnetization) only for DWs between two similar domains (AF-AF or FM-FM). Between a FM and an AF domain, however, we only consider the DW on the top FM block. Due to the extended FM stray field, the $B_z=0$ line is shifted towards the AF domain compared to the $m_z=0$ line. At 75 nm height above the surface, this shift amounts to several tens of nanometers according to our stray field simulation. From the profiles we can also see that above the DW, the IP field components $B_{x}$ and  $B_{y}$ have a local minimum and a maximum, respectively. In contrast to what has been observed in ultra-thin film \cite{Tetienne2015-rc,McLaughlin2023}, the profiles of the stray field intensity are not symmetric around the DW. We attribute this to the fact that we are imaging the DW at the top FM block  with a collective contribution of the full multilayered structure to the total stray field, \commt{in agreement with reported MFM profiles.\cite{Hellwig2003-prl,HELLWIG2007}} The DW-type can in principle be inferred by fitting the field profiles using an analytical expression of the stray field above the DW under the approximation of a thin sample \cite{Tetienne2015-rc}, i.e. a sample with magnetic thickness smaller than the sensor-sample distance. This approximation is not valid for the multilayered SAF investigated here since its magnetic thickness is comparable to the NV-sample distance. \commt{Previous micromagnetic simulations of multilayered structures demonstrated that FM domains are separated by Néel-type DWs at the top and bottom FM blocks with Bloch-type DWs  in the middle of the structure \cite{Salikhov2025}. Based on this, we could expect a similar DW-type configuration; however, a direct determination of the DW type would require additional high-resolution studies, such as iteratively refined micromagnetic simulations or complementary local magnetometry techniques (for instance, Lorentz transmission electron microscopy), which are beyond the scope of the present work.}

Our vector-field maps are in good agreement with  micromagnetic simulations.  Figures \ref{Fig:odmr}(d)-(f) show images obtained from micromagnetic simulation of the stray field components \commt{(see methods and section 11 in SI)}  at a distance of 75 nm from the sample surface, in the range of the estimated $d_{\text{NV}}$. The sensitivity of the extracted DW position to the experimental uncertainty in $d_{\text{NV}}$ can be evaluated by comparing the simulated stray field maps at heights of 36 nm and 75 nm. We find that varying the NV-sample distance within this range results in a change of approximately 3 nm in the absolute DW position, corresponding to a relative sensitivity of $\sim$ 8\% with respect to the total variation in $d_{\text{NV}}$. This value depends on the specific position of the profile used to extract the DW position; nevertheless, the variation is expected to remain of the same order even with a more rigorous statistical analysis. From the simulations we observed stray magnetic field of approximately $\pm$ 12 mT. The absolute values of the stray field are higher than the experimental values, suggesting that the magnetic parameters used to initially define the simulated structure need to be optimized as well as the determination of the NV-sample distance. However, the simulated and experimental stray-field maps are qualitatively in good agreement with the stray field values in the same order of magnitude, meaning that the remanent relaxed state achieved in the simulation can describe the magnetic texture of the multilayered SAF. Particularly, \commt{the granularity inducing a non-compensated magnetization on the AF domains is also reproduced in the simulated images. The FM domains along the stripe between the two AF domains are also well reproduced by the micromagnetic simulation. To visualize the domain structure in a 2D image, Figure \ref{Fig:odmr}(h) displays a transparent overlap of the simulated OOP magnetization of the  top and second FM blocks (see section 11 in SI)}. The FM domains around the DWs are formed due to an alternating left and right lateral shift $\delta$ between DWs in adjacent FM blocks, leading to a vertical alignment of the magnetic moments along the stacks and giving rise to a 3D spin texture as illustrated in the inset. This  lateral shift has been previously reported as driven  by the dipolar interaction against the AF-exchange interaction between the FM blocks with a periodic reversal of the FM cores to minimize the magnetostatic energy produced around the DWs, and with a \commt{constant DW-shift of $\delta \approx 20$ nm  along the FM stripe between AF domains for a multilayered structure with $X=6$ and $N=4$. \cite{Hellwig2003-prl}} In another study, Kiselev \emph{et al.} \cite{Kiselev2010} describe, \commt{within their micromagnetic model,} that the DW-shift configuration cannot be stabilized only by the interplay between the magnetostatic and the AF-exchange interaction. The stability depends also on DW pinning (grain boundaries, defects or inhomogeneities) that give rise to nucleation sites for one-dimensional FM stripes with fixed width (DW-shift). Remarkably, in the ODMR vector field maps, \commt{the periodic modulation of the stray field as a consequence of the DW-shift that reduce the magnetostatic energy} is quantitatively and spatially resolved in both the experimental and the simulated IP field maps in Figure \ref{Fig:odmr} (see black arrows). Additionally, the observed granularity or inhomogeneity can favor the stabilization of DW shifts through the DW pinning.  \commt{We found that the lateral DW-shift is non-uniform along the FM stripes}, it varies in the range  $\delta \approx 0-40$ nm (see section 11 in SI), generating FM domain cores with different nanometer sizes and wave-shaped DWs. 

\subsection{Sensing different components of the magnetic noise}

\begin{figure}[H]
    \centering
\includegraphics[width=6.6 true in]{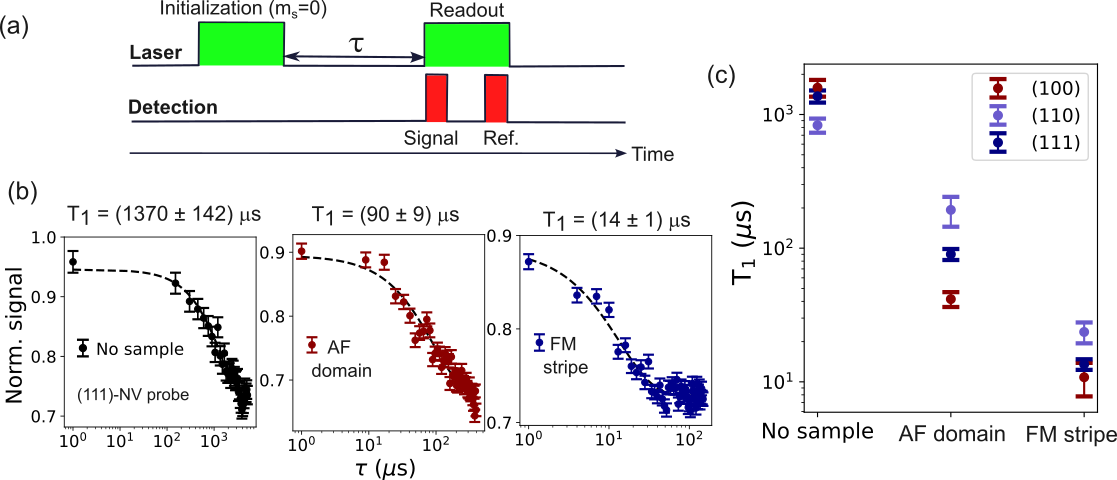}
\caption{\label{Fig:T1} \textbf{$T_1$ relaxometry and magnetic noise} (a) Pulse sequence for all-optical $T_1$ measurements. A green laser pulse of 3 $
\mu$s is sent to initially polarize the NV spin state in $m_s = 0$, after some time $\tau$ a second laser pulse is used for reading out the final NV spin state. The signal is read out during a time window of 200 ns (between the first 200-400 ns) of the second pulse and normalized by the signal during the last microsecond (reference signal).  (b) $T_1$ measurements performed on a (111)-NV probe with the tip withdrawn from the sample, and after positioning the tip on the AF domain and FM wall. (c) Comparison of $T_1$ measurements on diamond probes with different crystallographic orientations.}
\end{figure}

Finally, in order to understand the spin dynamics of the magnetic texture in the multilayered SAF, we performed $T_1$ relaxometry measurements to detect spin noise. $T_1$ is very sensitive to magnetic fluctuations perpendicular to the NV-axis and with frequencies near the NV resonance frequency \cite{Degen2017}. To investigate spin noise, we performed all-optical $T_1$ measurements \cite{Levine-2019} using the pulse sequences in  Figure \ref{Fig:T1}(a). We use a green laser pulse for initialization of the NV-spin states in $m_s=0$, wait for some time $\tau$ without laser illumination, and then a second green laser pulse is sent to read out the final spin state.   Figure \ref{Fig:T1}(b) shows the $T_1$ measurements on the (111)-oriented diamond tip. We first measure $T_1$ before approaching the sample surface, obtaining $T_1=(1370 \pm 142) \hspace{1mm}\mu$s. This value is expected for diamond tips with shallow NV (nominal implantation depth of 10 nm), where the relaxation time is mostly affected by paramagnetic impurities on the diamond surface \cite{Rosskopf2014}. After the sample approach and positioning the tip on the AF domain and then on the FM stripe, we observe a drop in $T_1$ by almost one and two orders of magnitude, respectively. We obtain $T_1=(90 \pm 9) \hspace{1mm}\mu$s when the tip is in the AF domain and even shorter  $T_1=(14 \pm 1) \hspace{1mm}\mu$s on the FM stripe. Recent calculations of the spin-wave spectrum in multilayered SAFs \cite{Jimenez-Bustamante2025-ub} predict GHz optical and acoustic spin-wave modes, with the dispersion relation and number of modes dependent on the thickness and number of FM blocks. Additionally, our results are in agreement with the results reported by Finco \emph{et al} \cite{Finco2021-ne} on the study of magnetic noise on AF DWs. Thus, we attribute these experimental observations to the presence of magnetic noise in the GHz range, most likely due to thermal magnons with frequencies near the NV spin transition frequency. $T_1$ is shorter on the FM stripes than on the AF domains, probably due to the confinement of thermal magnonic modes with a stronger noise or frequency that matches the electron spin resonance frequency of the NV at 2.87 GHz.%The periodic modulation of the demagnetizing field around the DWs can also accelerate the relaxation of the NV and reduce $T_1$.

Figure 5(c) shows a comparison of $T_1$ on NV diamond scanning probes with different crystallographic orientations \commt{(see section 12 in SI)}. The change in $T_1$ is essentially the same for the three diamond probes, indicating the existence of magnetic noise with components perpendicular to the respective NV-axis. These $T_1$ measurements suggest that the PL quenching observed on the FM stripes between AF domains during the acquisition of NV-PL maps could have contributions from both magnetic noise and spin-mixing. For the measurement with the (100)-NV probe, the dominant contribution to the NV-PL quenching comes probably from spin-mixing, as demonstrated by local ODMR measurements. A similar effect is expected for the (110)-NV tip. For the (111)-NV tip, the IP components resolved in the ODMR maps could in principle produce spin mixing. However, there is no evidence of spin-mixing from local ODMR measurements. Hence, the magnetic noise is the dominant contribution that produces quenching in the NV-PL images. Under this condition, it should be possible to observe an improvement in the PL contrast in the NV-PL images when the laser power is decreased below saturation, as observed on samples with AF DW  \cite{Finco2021-ne}. However, NV-PL quenching images recorded with different laser powers \commt{(section 13 in SI)} show no improvement in PL contrast, indicating that the field of the FM core along the DW could potentially play a role in the relation between the spin-polarization rate given by the laser power and the spin relaxation rate  defined by $T_1$.

In the model reported by Finco \emph{et al} \cite{Finco2021-ne}, the contrast in NV-PL quenching maps is described as a competition between the spin relaxation rate and the optical pumping rate, without considering the effect of the sample stray field. In this case, $m_s=\pm1$ states of the NV-probe are degenerate. In our multilayered SAF, in addition to the GHz magnetic noise, the FM cores generate a local magnetic field that varies spatially along the boundaries between AF domains. As a result, the $m_s=\pm1$ states are no longer degenerate and therefore have different occupation probabilities. This likely alters the initial conditions and symmetry assumed in the two-level model, which leads to different coupling conditions between the NV spin transitions and the magnetic noise. Consequently, the improvement of the PL contrast at reduced laser power is not observed in our mixed-order structure.

\section{Conclusion}

In conclusion, we present a detailed study for non-invasive and direct magnetic imaging of complex spin textures in a thick multilayered SAF with PMA using NV-SPM under ambient conditions, with vector-field sensitivity and nanoscale spatial resolution. We demonstrate that qualitative imaging with MFM-like contrast is achieved by applying an off-axis field \commt{to introduce a competition of the magnetic field components inducing NV-PL quenching}. Quantitative imaging is performed using an NV-probe with the NV-axis aligned with the sample magnetization, reducing possible spin-mixing effects even if the stray field in the sample is on the order of several mT. 

Quantitative vector-field maps combined with micromagnetic simulations allow us to explore interlayer coupling effects in the multilayered structure through direct magnetic imaging of DW shifts and the periodic modulation of the IP stray field components produced by the formation of nanometric FM cores around \commt{undulated DWs}. These results provide \commt{validation on an experimental and micromagnetic simulation level and further insights into existing models about the observed 3D spin texture}. Additionally, local $T_1$ relaxometry measurements reveal the presence of GHz-range magnetic noise on both the AF domains and the DWs with FM cores. 

Our findings contribute to the understanding of nanoscale magnetism in complex multilayered SAFs with coexisting AF and FM interactions. Quantitative direct magnetic imaging, along with spin noise sensing, offers new opportunities to investigate mixed spin phases, providing insights into domain and DW stability, interlayer coupling, and spin-wave modes. A proper description of these properties is fundamental for improving existing theoretical models and for designing advanced 3D magnetic architectures.

\section{Methods}

\subsection{Sample preparation and characterization }

The multilayered SAF was prepared at room temperature by dc magnetron sputtering deposition (ATC 2200 series sputter tool from AJA International, Inc.) with a base pressure of $8\cdot 10^{-6}$\,Pa, using Ar sputter gas with 0.4\,Pa pressure. Silicon substrates with a thermal 100\,nm SiO$_2$ layer were used, on which a 1.5\,nm thick Ta adhesion layer\cite{Ehrler2025}, 20\,nm Pt seed layer for promoting (111) texture\cite{Zeper1991}, and 3\,nm Pt cap against oxidation were deposited.

Hysteresis curve measurements were performed with a Quantum
Design MPMS3 superconducting interference device vibrating sample magnetometer (SQUID-VSM).

MFM measurements were performed using a Bruker Dimension Icon Atomic Force Microscope, employing probes model MESP with magnetic Co/Cr coating. All the MFM images were acquired in the lift imaging mode with tip-sample distance of 15 nm. 

\subsection{NV-SPM measurements}

Qualitative measurements (NV-PL maps and $T_1$)  were mostly performed in a new home-built setup, which combines tuning-fork AFM (attocube tip and sample positioner and scanner) with confocal optical microscopy. NV centers in the diamond probes are optically pumped with a 515 nm laser (Toptica, iBEAM-SMART-515-S) and readout by detecting the NV-PL signal after passing through a long-pass 650 nm filter (Thorlabs, FELH0650). The signal is collected by using dual single photon counting modules  (Excelitas, SPCM-AQRH-14) mounted in a Hanbury Brown and Twiss geometry for photon autocorrelation measurements. The excitation power of the green laser is  130-300 $\mu$W, measured after the objective lens (Olympus, MPLFLN100x) that focus (collect) the light on (from) the tip. We use Qzabre quantum scanning tips with sensitivity Q6-Q7 and nominal NV implantation depth of 10 nm. Quantitative measurements (ODMR and dual-iso B) were carried out in a commercial Quantum Scanning NV Microscope (QSM, Qzabre AG) equipped with an electromagnet to apply magnetic field in arbitrary directions. All the NV-SPM images were acquired using the non-contact amplitude modulation mode. The NV-sample distance, as well as, the orientation of the NV-axis in the laboratory frame are determined using the method described in reference \cite{Hingant_2015}, on a 1 $\mu$m-wide \commt{Ta/Pt/Co(0.6nm)/Pt}  wire with perpendicular magnetic anisotropy as a calibration sample (see SI for more details). 

\subsection{Micromagnetic simulations}

The multilayered SAF was modeled using Mumax$^3$ \cite{Vansteenkiste2014} with the interlayer exchange interaction implemented as a custom field \cite{Joos2023,Salikhov2025}. Four FM blocks with a thickness of 6.75\,nm, separated by a 0.75\,nm thick spacer, with $4\times4\times0.75\,$nm$^3$ cell size, and 8 times lateral periodic boundary conditions were used in the simulation. The simulation volume includes large lateral dimensions (2.048 $\mu$m × 2.048 $\mu$m) to match the experimental field of view and extends 75 nm above the sample surface to capture variations of the stray field within the estimated range for the NV-sample distance. To evaluate possible mesh-size effects in the description of the spin texture, a lateral mesh resolution of 2 nm was also tested, which resolves additional small-scale variations but does not significantly change the overall magnetization configuration. This high-resolution discretization of the mesh, however, increases both memory consumption and computation time. Therefore, the lateral cell size of 4 nm was chosen to balance the large simulation volume with manageable computation time.

The structure is initially defined as a perfect square AF domain with a sharp DW (see SI Figure S11). A relaxation and subsequent minimization was performed to let the magnetization evolve into the energetic minimum state. The saturation magnetization $M_\mathrm{s}= 833$\,kA/m from the SQUID-VSM measurement (see SI Figure S3) was used; other magnetic parameters were adopted from previous studies in similar multilayer structures: exchange stiffness $A=8.0$\,pJ/m \cite{Salikhov2021}, interlayer exchange coupling constant $A_\mathrm{IEC}=-0.7$\,mJ/m$^2$ \cite{Samad2021},  perpendicular anisotropy  $K_\mathrm{U}= 0.6$\,MJ/m$^3$ \cite{Koch2020}, average grain size 32\,nm with a normally distributed anisotropy direction variation among the grains \cite{Salikhov2022}. Second-order anisotropy terms and interfacial DMI were not considered in the model.

\section{Supporting Information}
Characterization of NV probes; determination of the NV-sample distance; MFM images and
magnetic hysteresis loop; additional NV-PL maps and ODMR measurements; estimation
of contrast in NV-PL maps exhibiting MFM-like contrast; iso-B images; vector magnetic
field reconstruction; close-up view of the ODMR contrast on AF domains; micromagnetic
simulation codes and magnetization simulations; T1 measurements; NV saturation curve and
NV-PL maps at different laser powers.

\begin{acknowledgement}

\commt{This work was supported by an Emmy Noether grant from the Deutsche Forschungsgemeinschaft (DFG, German Research Foundation), project No. 504973613. R.J.P.R. and A.S. acknowledge financial support by the Deutsche Forschungsgemeinschaft (DFG, German Research Foundation) through the Würzburg–Dresden Cluster of Excellence ctd.qmat – Complexity, Topology and Dynamics in Quantum Matter (EXC 2147, project-id 390858490). Furthermore, this work was supported by the DFG through the project No. 514946929 at Chemnitz University of Technology.}

\end{acknowledgement}


\bibliography{biblio}

%\includepdf[pages=-]{Supp_Info.pdf}

\end{document}